\documentclass[letterpaper,12pt]{article}   
\usepackage{osajnl2} 
\usepackage{epsfig}
\begin{document}


\title{Photon-number statistics with Silicon photomultipliers}
\author{Marco Ramilli,$^1$ Alessia Allevi,$^2$ Valery Chmill,$^1$ Maria Bondani,$^{3,4,*}$\\
Massimo Caccia,$^1$ and Alessandra Andreoni$^{1,4}$}
\address{$^1$Dipartimento di Fisica e
Matematica, Universit\`a degli Studi dell'Insubria\\
I-22100, Como, Italy}
\address{$^2$CNISM, U.d.R.  Milano Universit\`a, I-20133, Milano, Italy}
\address{$^3$National Laboratory for Ultrafast and Ultraintense
Optical Science - CNR-INFM,  I-22100, Como, Italy}
\address{$^2$CNISM, U.d.R.  Milano Universit\`a, I-20133, Milano, Italy}
\address{$^4$CNISM, U.d.R.  Como, I-22100, Como, Italy}
\address{$^*$Corresponding author: maria.bondani@uninsubria.it}

\begin{abstract}
We present a description of the operation of a multi-pixel detector in the presence of
non-negligible dark-count and cross-talk effects. We apply the model to devise
self-consistent calibration strategies to be performed on the very light under
investigation.
\end{abstract}
\maketitle

\section{Introduction}\label{sec:intro}
In the last decade many efforts have been devoted to the development and characterization
of photon-number resolving detectors. The motivation stems from the variety of fields of
applications: as an example, this kind of devices has already been used in particle
physics experiments \cite{Kanaya,Buzhan,Dolgo,Haba,Korpar,Take}, PET systems
\cite{Dolgo,Mohers}, biomedical research \cite{Grigoriev,Gomi,Cappellini,Fukuda,Risigo}
and atmospheric pollution measurements \cite{Kedar,Arnon}. Moreover, there are other
interesting fields of application in which photon-counting detectors could be exploited,
such as LIDAR (light detection and ranging), laser scanning microscopy, and fluorescence
correlation spectroscopy \cite{Hamamatsu}. In addition, this class of detectors play a
key role in quantum optics as it offers the possibility to characterize nonclassical
multiphoton states. In this field they can be considered as a valid alternative to
optical homodyne tomography \cite{Lvovsky09} as they can be employed in direct detection
schemes \cite{OL09}.

Several concepts and technologies have been proposed that lead to the development of
detectors such as visible-light photon counters (VLPC) \cite{Yamamoto99}, superconductive
transition edge sensors (TES) \cite{Lita}, time-multiplexed detectors
\cite{Achilles,Laiho, Ondrej}, hybrid photodetectors (HPD) \cite{JMO,ASL} and Silicon
photomultipliers (SiPM) \cite{Akindinov}. Irrespective of the concept and design
features, these detectors may in general be classified in terms of photon detection
efficiency, spectral response, time development of the signal, dead time, and notable
photon number resolving capability.  As of today, the ideal detector has yet to appear
and the optimal choice is application specific.

This paper focuses on SiPMs, detectors featuring unique characteristics that are achieved
by a rapidly evolving technology. Silicon photomultipliers consist of a high density (by
now limited to $\sim 2000$ cells/mm$^{2}$) matrix of diodes with a common output. Each
diode (or cell) is operated in a limited Geiger-Mueller (GM) regime, in order to achieve
gains at the level of $10^6$. Quenching mechanisms are implemented to avoid establishing
self-sustaining discharges. These detectors are sensitive to single photons triggering GM
avalanches and can be endowed with a dynamic range well above 100~photons/burst. The
photon detection efficiency (PDE) depends on the sensor design and specification, but it
may well exceed 60\%. Moreover, SiPM are genuine photon-number resolving detectors in
that they measure light intensity simply by the number of fired diodes. Compactness,
robustness, low cost, low operating voltage, and power consumption are also added values
against traditional photodetectors. On the other hand, SiPMs are affected by significant
dark count rates (DCR), associated to cells fired by thermally generated charge carriers.
Moreover, the GM avalanche development is known to be associated to the generation of
photons \cite{COVA}, which may in turn trigger  secondary avalanches and result in
relevant cross-talk. Whether DCR and cross-talk may be directly measured, it is clear
that they are folded in the detector response to any signal and need to be modelled to
properly assess the statistical properties of the light field being investigated. This
paper reports the experimental validation of two models, on the way to a self-consistent
characterization of the SiPM response.

\section{Experimental set-up}\label{sec:setup}

\begin{figure}[tb]
\begin{center}
\includegraphics[width=0.8\textwidth]{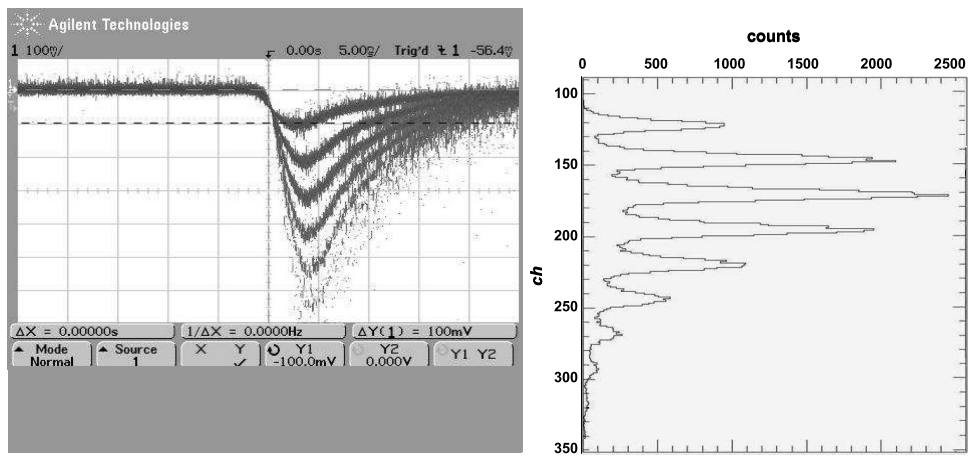}
\caption{Left panel: Output of a SiPM Hamamatsu  MPPC S10362-11-100C detecting a weak light
field as displayed by a 100 MHZ oscilloscope. Right panel: histogram of the corresponding spectrum.}
\label{pic:peaks}
\end{center}
\end{figure}

\begin{figure}[tb]
\begin{center}
\includegraphics[width=1\textwidth]{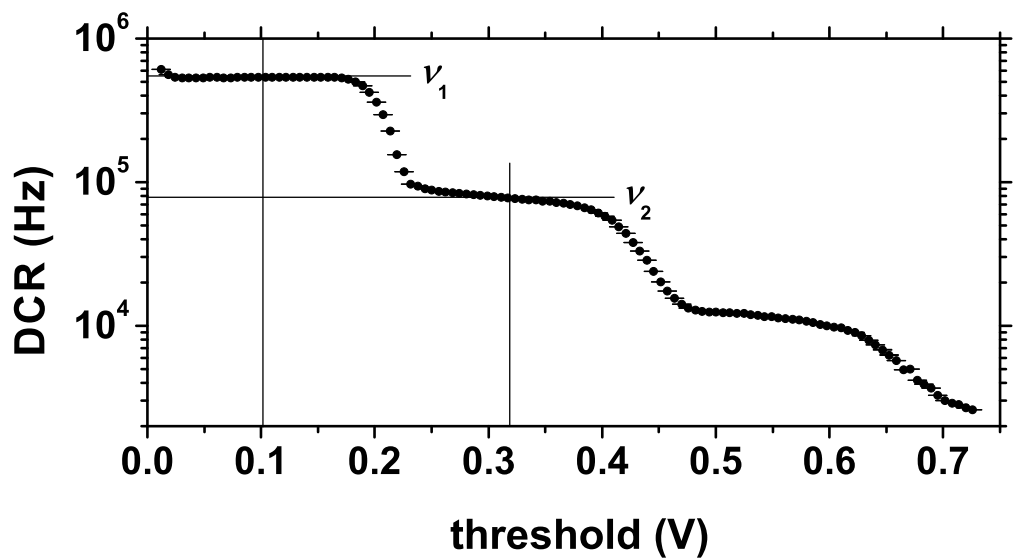}
\end{center}
\caption{DCR threshold scan of the SiPM performed at room temperature with no impinging light. }\label{pic:staircase}
\end{figure}

\begin{figure}[h]
\begin{center}{
 \includegraphics[angle=0,width=1\textwidth]{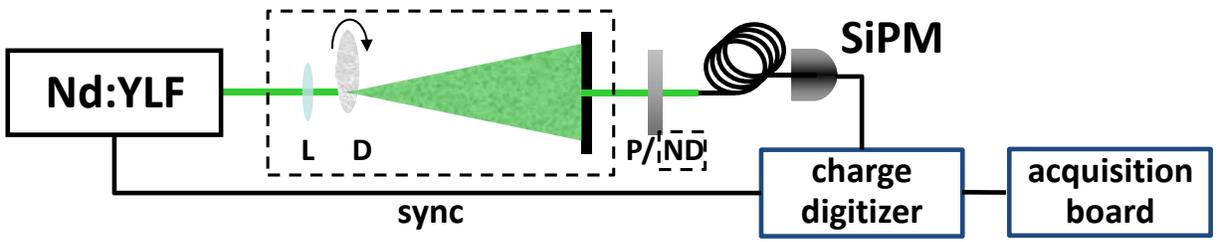} }
\caption{(Color online) Experimental setup. Nd:YLF: laser source, P: polaroid, ND:
variable neutral density filter, SiPM: detector. The components in the dashed boxes
are inserted to produce the pseudo-thermal field.} \label{f:setup}
\end{center}
\end{figure}

The detector response to a weak light field is shown in Fig.~\ref{pic:peaks}(a),
featuring the sensor output signal after a high-bandwidth amplifier with a gain of 50.
The different bands in the image, obtained in persistency mode, correspond to samples
with different numbers of triggered cells, i.e. different numbers of detected photons.
The photon-number resolving properties are also clear in Fig.~\ref{pic:peaks}(b) that
shows the corresponding spectrum as obtained by digitizing the amplified current pulse
from the detector over a well defined time window, synchronized to the light pulse. These
data and the results reported in the following were obtained with a SiPM produced by
Hamamatsu Photonics \cite{Hamamatsu}; more specifically, the main features of the
detector are reported in Table \ref{tab:MPPC}. We have decided to test a detector with
$10^2$ cells in order to keep DCR and cross-talk at a reasonable level and to maximize
PDE. The signal was integrated by a charge digitizer V792 produced by Caen \cite{CAEN},
with a 12-bit resolution over 400~pC range; the signal was typically integrated over a
200~ns long time window. The DCR and cross-talk can be directly obtained by measuring the
frequency of the pulses from the sensor, with no illumination, above a well defined
threshold. The result of a threshold scan of DCR is displayed in
Fig.~\ref{pic:staircase}, clearly showing that the single cell avalanches stochastically
triggered by thermally generated carriers can induce, by optical cross-talk, a cascade
effect resulting in a multiple cell spurious event. Thus optical cross-talk can be
measured using:
\begin{equation}
X_{talk} = \frac{\nu_{2}}{\nu_{1}}
\end{equation}
where $\nu_{1}$ is the the dark count rate measured by setting a threshold at a value
lower than the peak amplitude of the signal corresponding to a single avalanche and
$\nu_{2}$ is the count rate obtained when the threshold is set beyond the single cell
signals (see Figure~\ref{pic:staircase}). The experimental results reported below were
obtained illuminating the sensor with a frequency-doubled Nd:YLF mode-locked laser
amplified at 500 Hz (High Q Laser Production) that provides linearly polarized pulses of
$\sim$5.4~ps duration at 523 nm wavelength. Two series of measurements were performed,
the first one directly on the coherent laser output and the second one on the
pseudo-thermal light obtained by passing the laser through a rotating ground-glass
diffuser (D in Fig.~\ref{f:setup}) \cite{Arecchi1965}. The light to be measured was
delivered to the sensor by a multimode optical fiber (1~mm core diameter). The signal
digitization was synchronized to the laser pulse.

%
\section{Detector response modelling}\label{sec:teo}

The response of an ideal detector to a light field can be described in a simple way as  a
bernoullian process:
\begin{equation}
 B_{m,n}(\eta)=\left(
 \begin{array}{c}n\\m\end{array}\right)
 \eta^m (1-\eta)^{n-m}\ ,\label{eq:bern}
\end{equation}
being $n$ the number of impinging photons over the integration time, $m$ the number of
detected photons and $\eta<1$ the photon-detection efficiency. Actually, $\eta$ is a
single parameter quantifying detector effects and losses (intentional or accidental)
which can be tracked to the optical system. As far as SiPMs are concerned, detector
effects are due to the quantum efficiency, the fill factor and the avalanche triggering
probability, namely the probability for a charge carrier to develop a Geiger-Mueller
quenched discharge \cite{mckay,oldham}. As a consequence, the distribution
$P_{m,\mathrm{el}}$ of the number of detected photons, that is the GM avalanches actually
corresponding to a detected photon has to be linked to the distribution $
P_{n,\mathrm{ph}}$ of the number of photons in the light under measurement by
\cite{mandel1995, agliati2005, zambra2004}:
\begin{equation}
 P_{m,\mathrm{el}}=\sum_{n=m}^{\infty}B_{m,n}(\eta) P_{n,\mathrm{ph}}\ .\label{eq:phel}
\end{equation}
It can be demonstrated \cite{casini,ASL} that for a combination of classical light states
the statistics is preserved by the primary detection process. This simple description has
to be further developed to link $P_{m,\mathrm{el}}$ to the probability density
distribution of the GM avalanches of any origin. First we must take into account spurious
hits and cross-talk effects, not negligible in the detectors being studied. The dark
count rate results in a poissonian process which can be described as:

\begin{equation}
 P_{m,\mathrm{dc}}=\overline{m}_{\mathrm{dc}}^m/m! \exp(-\overline{m}_{\mathrm{dc}})\ , \label{eq:DC}
\end{equation}
where $\overline{m}_{\mathrm{dc}}$ is the mean number of dark counts  during the gate
window (or integration time) and
$\sigma_{m,\mathrm{dc}}^{(2)}=\sigma_{m,\mathrm{dc}}^{(3)}=\overline{m}_{\mathrm{dc}}$.

As a consequence, the statistics of the recorded pulses may be described  as:
\begin{equation}
 P_{m,\mathrm{el+dc}}=\sum_{i=0}^{m} P_{i,\mathrm{dc}} P_{m-i,\mathrm{el}}\ ,\label{eq:phelDCR}
\end{equation}
obviously shifting the mean value to
$\overline{m}_{\mathrm{el+dc}}=\overline{m}_{\mathrm{el}}+\overline{m}_{\mathrm{dc}}$,
with an increased variance and third-order central moment
$\sigma^{(2)}_{m,\mathrm{el+dc}}=\sigma^{(2)}_{m,\mathrm{el}}+\sigma^{(2)}_{m,\mathrm{dc}}=
\sigma^{(2)}_{m,\mathrm{el}}+\overline{m}_{\mathrm{dc}}$ and
$\sigma^{(3)}_{m,\mathrm{el+dc}}=\sigma^{(3)}_{m,\mathrm{el}}+\sigma^{(3)}_{m,\mathrm{dc}}=
\sigma^{(3)}_{m,\mathrm{el}}+\overline{m}_{\mathrm{dc}}$.

As a further step, cross-talk effects shall be taken into account. Cross-talk is a
genuine cascade phenomenon that can be described at first order as \cite{Silberberg}

\begin{equation}
 C_{k,l}(\epsilon)=\left(
 \begin{array}{c}l\\k-l\end{array}\right)
 \epsilon^{k-l} (1-\epsilon)^{2l-k}\ .\label{eq:cross}
\end{equation}
being $\epsilon$ the (constant) probability that the GM avalanche of a cell triggers a
second cell, $l$ the number of dark counts and photo-triggered avalanches and $k$ the
actual light signal amplitude. Within this first-order approximation, the actual sensor
response is described by
\begin{equation}
 P_{k,\mathrm{cross}}=\sum_{m=0}^{k} C_{k,m}(\epsilon)P_{m,\mathrm{el+dc}}\ ,\label{eq:crossDC}
\end{equation}
characterized by
$\overline{k}_{\mathrm{cross}}=(1+\epsilon)\overline{m}_{\mathrm{el+dc}}$,
$\sigma^{(2)}_{k,\mathrm{cross}}=(1+\epsilon)^2\sigma^{(2)}_{m,\mathrm{el+dc}}+\epsilon(1-\epsilon)\overline{m}_{m,\mathrm{el+dc}}$
and $\sigma^{(3)}_{k,\mathrm{cross}}=(1+\epsilon)^3\sigma^{(3)}_{m,\mathrm{el+dc}}+
3\epsilon(1-\epsilon^2)\sigma^{(2)}_{m,\mathrm{el+dc}}+\epsilon(1-3\epsilon+2\epsilon^2)\overline{m}_{m,\mathrm{el+dc}}$.
In the following we refer to this analytical model as \emph{Model~I}. We also consider a
better refined model (\emph{vide infra}, \emph{Model~II}) offering, in principle, an
extended range of application, but paying the price of being limited to a numerical
rather than analytical solution. Irrespective of the model, the amplification and
digitization processes that produces the output $x$ can simply be described as a
multiplicative parameter $\gamma$:
\begin{equation}
 P_{x,\mathrm{out}}=\gamma P_{\gamma k,\mathrm{cross}}\ ,\label{eq:out}
\end{equation}
conveniently scaling the momenta as
$\overline{x}_{\mathrm{out}}=\gamma\overline{k}_{\mathrm{cross}}$,
$\sigma^{(2)}_{x,\mathrm{out}}=\gamma^2\sigma^{(2)}_{k,\mathrm{cross}}$ and
$\sigma^{(3)}_{x,\mathrm{out}}=\gamma^3\sigma^{(3)}_{k,\mathrm{cross}}$.
\subsection{\emph{Model~I}: an analytical evaluation of the second and third order momenta}

This approach extends the method presented in  \cite{JMO}  to detectors with a
significant dark count rate and first order cross-talk effects. Experimentally, it is
based on the detection of a light field performed by varying, in a controlled way, the
optical losses, i.e. $\eta$, with detector parameters presumed to be constant throughout
the $\eta$ scan. The second-order momentum of the recorded pulse distribution
$P_{x,\mathrm{out}}$ can be used to evaluate the Fano factor:
\begin{eqnarray}
 F_{x,\mathrm{out}}&=&\frac{\sigma^{(2)}_{x,\mathrm{out}}}{\overline{x}_{\mathrm{out}}}=\gamma \frac{\sigma^{(2)}_{k,\mathrm{cross}}}{\overline{k}_{\mathrm{cross}}}\nonumber\\
 &=&\gamma(1+\epsilon)\frac{\sigma^{(2)}_{m,\mathrm{el+dc}}}{\overline{m}_{\mathrm{el+dc}}}+ \gamma\frac{\epsilon(1-\epsilon)}{1+\epsilon}\nonumber\\
 &=& \frac{Q_{\mathrm{el+dc}}}{\overline{m}_{\mathrm{el+dc}}} \overline{x}_{\mathrm{out}} + \gamma\frac{1+3\epsilon}{1+\epsilon}\ ,\label{eq:outF}
\end{eqnarray}
where $Q_{\mathrm{el+dc}}=\sigma^{(2)}_{m,\mathrm{el+dc}}/\overline{m}_\mathrm{el+dc}-1$
is the Mandel factor of the primary charges. Note that, due to dark-counts, the
coefficient of ${\overline{x}}_{\mathrm{out}}$ in Eq.~(\ref{eq:outF}) cannot be written
as $Q_{\mathrm{ph}}/\overline{n}$ \cite{JMO, Aa}, that is, the coefficient
$Q_{\mathrm{el+dc}}/\overline{m}_{\mathrm{el+dc}}$ does not only depend on the light
field to be measured. Similarly we can calculate a sort of symmetry parameter
\begin{eqnarray}
 S_{x,\mathrm{out}}&=&\frac{\sigma^{(3)}_{x,\mathrm{out}}}{\overline{x}_{\mathrm{out}}}=\gamma^2 \frac{\sigma^{(3)}_{k,\mathrm{cross}}}{\overline{k}_{\mathrm{cross}}}\nonumber\\
 &=&\gamma^2(1+\epsilon)^2\frac{\sigma^{(3)}_{m,\mathrm{el+dc}}}{\overline{m}_{\mathrm{el+dc}}}+ 3\gamma^2\epsilon(1-\epsilon)\frac{\sigma^{(2)}_{m,\mathrm{el+dc}}}{\overline{m}_{\mathrm{el+dc}}}+ \gamma^2\frac{\epsilon(1-3\epsilon+2\epsilon^2)}{1+\epsilon}\nonumber\\
 &=& \frac{Q_{s,\mathrm{el+dc}}-3Q_{\mathrm{el+dc}}}{\overline{m}_{\mathrm{el+dc}}^2} \overline{x}_{\mathrm{out}}^2 + \gamma\frac{1+3\epsilon}{1+\epsilon}\frac{Q_{\mathrm{el+dc}}}{\overline{m}_{\mathrm{el+dc}}} \overline{x}_{\mathrm{out}} + \gamma^2\frac{1+7\epsilon}{1+\epsilon}\ ,\label{eq:outS}
\end{eqnarray}
where
$Q_{s,\mathrm{el+dc}}=\sigma^{(3)}_{m,\mathrm{el+dc}}/\overline{m}_\mathrm{el+dc}-1$.
Note that in the presence of dark-counts both coefficients
$(Q_{s,\mathrm{el+dc}}-3Q_{\mathrm{el+dc}})/\overline{m}_{\mathrm{el+dc}}^2$ and
$Q_{\mathrm{el+dc}}/\overline{m}_{\mathrm{el+dc}}$ are no-more independent of the light
under measurement \cite{ASL}. We will see how this modifies the results in the following
examples involving classical light states for which the statistics of detected photons is
the same as that of photons.

A validation and a comparison of the proposed model was performed by sampling coherent
and multi-thermal light fields, where the proposed model can be specified as follows.

\subsubsection{Sampling a coherent light field}

For a coherent field, the photon-number distribution is given by
\begin{equation}
 P_{n,\mathrm{ph}}  = \frac{|\alpha|^{2n}}{n!} e^{-|\alpha|^2}\ ,\label{eq:poiss}
\end{equation}
for which $\overline{n}=\sigma_{n}^{(2)}=\sigma_{n}^{(3)}=|\alpha|^2$. Remembering that
the distribution of dark-counts is also poissonian, according to Eq.s~(\ref{eq:outF}) and
(\ref{eq:outS}), we have:
\begin{eqnarray}
 &&\frac{Q_{\mathrm{el+dc}}}{\overline{m}_{\mathrm{el+dc}}} =  \frac{1}{\overline{m}_{\mathrm{el+dc}}}\left[\frac{\sigma^{(2)}_{m,\mathrm{el+dc}}}
 {\overline{m}_\mathrm{el+dc}} -1\right]=\frac{\sigma^{(2)}_{m,\mathrm{el}}+\sigma^{(2)}_{m,\mathrm{dc}}-(\overline{m}_\mathrm{el}+\overline{m}_\mathrm{dc})}
 {(\overline{m}_\mathrm{el}+\overline{m}_\mathrm{dc})^2} = 0\nonumber\\
 &&\frac{Q_{s,\mathrm{el+dc}}-3Q_{\mathrm{el+dc}}}{\overline{m}_{\mathrm{el+dc}}^2} = \frac{1}{\overline{m}_{\mathrm{el+dc}}^2}\left[\frac{\sigma^{(3)}_{m,\mathrm{el+dc}}}
 {\overline{m}_\mathrm{el+dc}} -1\right] =\frac{\sigma^{(3)}_{m,\mathrm{el}}+\sigma^{(3)}_{m,\mathrm{dc}}-(\overline{m}_\mathrm{el}+\overline{m}_\mathrm{dc})}
 {(\overline{m}_\mathrm{el}+\overline{m}_\mathrm{dc})^3}= 0\ .\nonumber\label{eq:Qcoh}
\end{eqnarray}
We thus obtain from Eq.s~(\ref{eq:outF}) and (\ref{eq:outS})
\begin{eqnarray}
  F_{x,\mathrm{out}} &=& \gamma\frac{1+3\epsilon}{1+\epsilon}\label{eq:fanoCOH}\\
  S_{x,\mathrm{out}} &=& \gamma^2\frac{1+7\epsilon}{1+\epsilon}\label{eq:skewCOH}
\end{eqnarray}
at any mean value $\overline{x}_{\mathrm{out}}$.

The results in Eq.s~(\ref{eq:fanoCOH}) and (\ref{eq:skewCOH}) show that, if we measure a
coherent field at different values of ${\overline{x}}_{\mathrm{out}}$ by varying the
overall detection efficiency $\eta$ through an attenuator \cite{JMO}, we can obtain the
values of $\gamma$, the detector gain, and $\epsilon$, the probability of cross-talk, by
fitting the experimental values of $F_{x,\mathrm{out}}$ and $S_{x,\mathrm{out}}$. The
$\gamma$-value is obtained in units of $x$ per detected photon. Of course, no information
can be obtained from Eq.s~(\ref{eq:fanoCOH}) and (\ref{eq:skewCOH}) on the amount of
dark-counts as far as the statistics is poissonian for both light and dark-counts. Note
that, as the relation between $\gamma$ and $\epsilon$ is quadratic, we have in general
two sets of possible values for each fit. The choice of one of the sets is made by
comparing the value of $\gamma$ with the interspacing between the peaks of the
pulse-height spectrum.

\subsubsection{Sampling a multi-mode thermal light field}

The photon-number distribution of a field made of $\mu$ independent thermal
modes each containing $N_{th}/\mu$ mean photons \cite{mandel1995} is given by
\begin{equation}
 P_{n,\mathrm{ph}} = \frac{\left(n +\mu-1\right)!}
 {n!\left(\mu - 1 \right)! \left(N_{th}/\mu+1 \right)^{\mu} \left(\mu/N_{th}+1 \right)^{n}}\ , \label{eq:multit}
\end{equation}
for which $\overline{n}=N_{th}$, $\sigma^{(2)}_{n}= N_{th}\left(N_{th}/\mu+1 \right)$ and $\sigma^{(3)}_{n}= N_{th}\left(N_{th}/\mu+1 \right)\left(2N_{th}/\mu+1 \right)$. In this case we have
\begin{eqnarray}
 \frac{Q_{\mathrm{el+dc}}}{\overline{m}_{\mathrm{el+dc}}} &=&  \frac{1}{\overline{m}_{\mathrm{el+dc}}} \left[\frac{\sigma^{(2)}_{m,\mathrm{el+dc}}}{\overline{m}_\mathrm{el+dc}} -1\right]=\frac{\sigma^{(2)}_{m,\mathrm{el}}+\sigma^{(2)}_{m,\mathrm{dc}}-(\overline{m}_\mathrm{el}+\overline{m}_\mathrm{dc})}
 {(\overline{m}_\mathrm{el}+\overline{m}_\mathrm{dc})^2} \nonumber\\ &=&\frac{\overline{m}_\mathrm{el}(\overline{m}_\mathrm{el}/\mu+1)+\overline{m}_\mathrm{dc}
 -(\overline{m}_\mathrm{el}+\overline{m}_\mathrm{dc})}
 {(\overline{m}_\mathrm{el}+\overline{m}_\mathrm{dc})^2} = \frac{1}
 {\mu}\frac{\overline{m}_\mathrm{el}^2}
 {(\overline{m}_\mathrm{el}+\overline{m}_\mathrm{dc})^2}\nonumber\\
 \frac{Q_{s,\mathrm{el+dc}}-3Q_{\mathrm{el+dc}}}{\overline{m}_{\mathrm{el+dc}}^2} &=& \frac{1}{\overline{m}_{\mathrm{el+dc}}^2}   \left[\frac{\sigma^{(3)}_{m,\mathrm{el+dc}}}{\overline{m}_\mathrm{el+dc}} -1-3\frac{\sigma^{(2)}_{m,\mathrm{el+dc}}}{\overline{m}_\mathrm{el+dc}} +3 \right]\nonumber\\
 &=&\frac{\sigma^{(3)}_{m,\mathrm{el}}+\sigma^{(3)}_{m,\mathrm{dc}}-3\sigma^{(2)}_{m,\mathrm{el}}-3\sigma^{(2)}_{m,\mathrm{dc}}+2(\overline{m}_\mathrm{el}+\overline{m}_\mathrm{dc})}
 {(\overline{m}_\mathrm{el}+\overline{m}_\mathrm{dc})^3}\nonumber\\ &=&\frac{\overline{m}_\mathrm{el}(\overline{m}_\mathrm{el}/\mu+1)
 (2\overline{m}_\mathrm{el}/\mu+1)+\overline{m}_\mathrm{dc}-3\overline{m}_\mathrm{el}(\overline{m}_\mathrm{el}/\mu+1)-3\overline{m}_\mathrm{dc}
 +2(\overline{m}_\mathrm{el}+\overline{m}_\mathrm{dc})}
 {(\overline{m}_\mathrm{el}+\overline{m}_\mathrm{dc})^3}\nonumber\\
 &=& \frac{2}
 {\mu^2}\frac{\overline{m}_\mathrm{el}^3}
 {(\overline{m}_\mathrm{el}+\overline{m}_\mathrm{dc})^3}\nonumber\ .\nonumber\label{eq:Qmulti}
\end{eqnarray}
We now observe that the measured output mean value can be written as
$\overline{x}_{\mathrm{out}}=\gamma(1+\epsilon)(\overline{m}_\mathrm{el}+\overline{m}_\mathrm{dc})
\equiv \overline{x}_\mathrm{ph}+\overline{x}_\mathrm{dc}$, where only
$\overline{x}_\mathrm{ph}$ undergoes attenuation during the experimental procedure while
$\overline{x}_\mathrm{dc}$ remains constant. We thus rewrite Eq.s~(\ref{eq:outF}) and
(\ref{eq:outS}) as
\begin{eqnarray}
  F_{x,\mathrm{out}} &=& \frac{1}{\mu}\left(1-\frac{\overline{x}_{\mathrm{dc}}}{\overline{x}_{\mathrm{out}}}\right)^2 \overline{x}_{\mathrm{out}}+ \gamma\frac{1+3\epsilon}{1+\epsilon}\label{eq:FANOmultit}\\
  S_{x,\mathrm{out}} &=& \frac{2}{\mu^2}\left(1-\frac{\overline{x}_{\mathrm{dc}}}{\overline{x}_{\mathrm{out}}}\right)^3 \overline{x}_{\mathrm{out}}^2 +\frac{3}{\mu}\gamma\frac{1+3\epsilon}{1+\epsilon}\left(1-\frac{\overline{x}_{\mathrm{dc}}}{\overline{x}_{\mathrm{out}}}\right)^2 \overline{x}_{\mathrm{out}} +\gamma^2\frac{1+7\epsilon}{1+\epsilon}\ .\label{eq:SKEWmultit}
\end{eqnarray}
By measuring our multi-mode thermal field at different values of
$\overline{x}_{\mathrm{out}}$, and plotting $F_{x,\mathrm{out}}$ and $S_{x,\mathrm{out}}$
we can obtain an estimation of the parameters $\mu$, $\overline{x}_{\mathrm{dc}}$,
$\epsilon$, $\gamma$ from a self-consistent procedure and without need of an independent
calibration. Again $\gamma$ is in units of $x$ per detected photon and the relation
between $\gamma$ and $\epsilon$ is quadratic (see above).


\subsection{\emph{Model~II}: a numerical evaluation based on the photon-number resolving properties of SiPM }
The above mentioned self-consistent method is very powerful, but requires the acquisition
of several histograms at varying $\eta$, which could not always be easy to perform, or
possible at all, in many practical applications: from this point of view, the possibility
to analyze each spectrum independently looks complementary to the self-consistent
approach. We performed this analysis with a two-step procedure:
\begin{itemize}
\item[-]{we measured the areas of the spectrum peaks, thus obtaining an estimation of the number of counts per peak;}
\item[-]{we fitted the obtained data points with a theoretical function, which takes into account the statistics of light, detection and all deviations of the detectors from ideality, such as DCR and cross-talk effects.}
\end{itemize}
To evaluate the area of each peak, we performed a multi-peak fit of the spectrum histogram, modelling each peak with a Gauss-Hermite function~\cite{gausshermite}:
\begin{equation}
\mbox{GH}\; = \; Ne^{-w^2/2}\:\left[1+\:^3hH_3(w)\:+\:^4hH_4(w)\right],\label{eq:gh}
\end{equation}
where
\begin{equation}
w=\frac{x-\bar{x}}{\sigma}\label{eq:w}
\end{equation}
and $N$ is a normalization factor, $\bar{x}$ is the peak position and $\sigma$ is the
variance of the gaussian function; $H_3(w)$ and $H_4(w)$ are the third and the fourth
normalized hermite polynomials and their contribution gives the asymmetry of the peak
shape, whose entity is regulated by the pre-factors $^3h$ and $^4h$, with values in the
range $[-1,1]$. The global fit function of the spectrum is a sum of as many Gauss-Hermite
function as the number of resolved peaks.\\
The choice of the GH-function in Eq.~(\ref{eq:gh}) allows us to calculate the area $A_n$
of the $n$-th peak in a very straightfoward way, simply by the relation
\begin{equation}
A_n = N_n\sigma_n(\sqrt{2\pi}+^4h_n).\label{eq:area}
\end{equation}
The error $\sigma_{A_{n}}$ on the obtained value is calculated by propagating the errors
on the fit parameters.\\
This analysis is also useful in order to calculate the system gain $\gamma$: in fact,
from the fitted values of the peak positions $\bar{x}_n$  we can calculate the
peak-to-peak distance $\Delta$ for all the resolved peaks:
\begin{equation}
\Delta_{n,n+1}= \bar{x}_{n+1}-\bar{x}_n.\label{eq:delta}
\end{equation}
The error $\sigma_{\Delta_{n,n+1}}$ associated to this value is once again obtained by
propagating the fit errors of the peak position values; furthermore, to estimate $\gamma$
a weighted average on all the peak-to-peak values obtained from the analyzed histogram is
performed.\\
As for the theoretical function, the effect of detection, DCR and amplification is
modelled as described in previous sections (see Eq.s~(\ref{eq:bern})-(\ref{eq:phelDCR})
and Eq.~(\ref{eq:out})).\\
The effect of cross-talk, is described by using a bernoullian process, in a way analogue
to what has been done with function $C_{k,m}(\epsilon)$ of Eq.~(\ref{eq:cross}). However,
as cross-talk process is intrinsically a cascade phenomenon, its contribution has been
calculated by adding higher order effects:
\begin{equation}
P_{k,cross}=\sum_{m=0}^{k}\:\sum_{n=0}^m\:\sum_{j=0}^n\,P_{k-m-n-j,el+dc}\,B_{m,k-m-n-j}(\epsilon)\,B_{n,m}(\epsilon)\,B_{j,n}(\epsilon);\label{eq:ho-cross}
\end{equation}
where terms like $B_{j,n}(\epsilon)$ stand for the bernoullian distribution
\begin{equation}
 B_{j,n}(\epsilon)=\left(
 \begin{array}{c}n\\j\end{array}\right)
 \epsilon^{j} (1-\epsilon)^{n-j}\ .
\end{equation}
Such a higher order expansion is not trivial to be achieved by the self-consistent
approach of \emph{Model~I}, in which an explicit analytic expression of
$P_{x,\mathrm{out}}$ is needed in order to calculate its momenta. Here, as all the
elements of interest ($\overline{m}_{el}$, $\overline{m}_{dc}$, $\epsilon$, the number of
modes $\mu$) will be obtained as fit parameters, this is not necessary and therefore
$P_{x,\mathrm{out}}$ can be just numerically calculated as the fitting function.\\
The major limitation of this approach is obvious: as all the information on the
statistics of the system is obtained from the peak areas, this method can only be applied
to peak-resolving histograms with a number of peaks greater than the number of free
parameters of the fitting function, which, in the present analysis, can rise up to five.

\section{Experimental results}

\subsection{Coherent light}

\begin{figure}[h]
\begin{center}{
 \includegraphics[angle=0,width=0.7\textwidth]{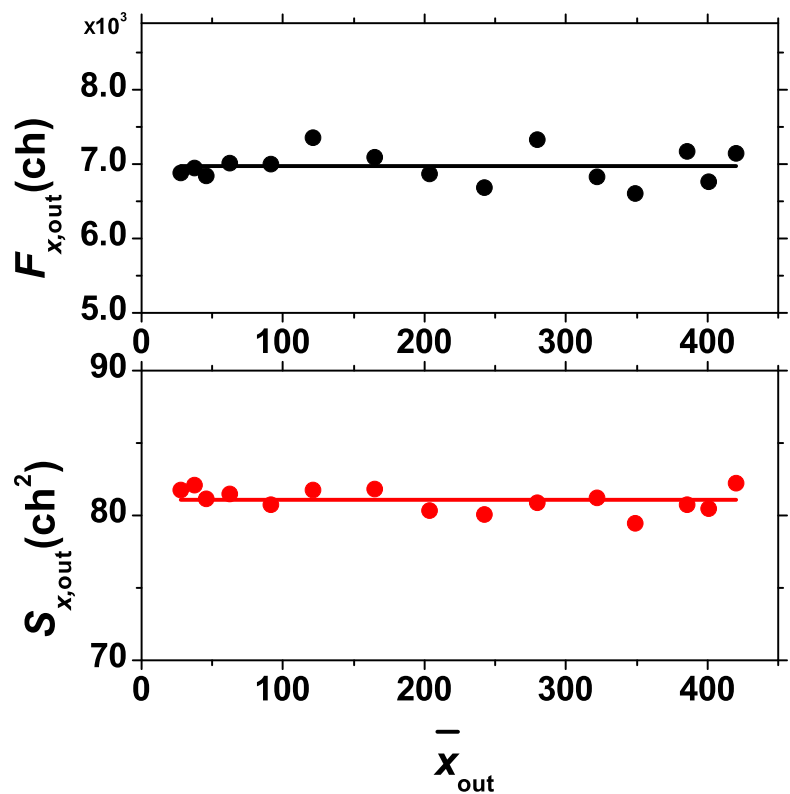} }
\caption{(Color online) Plot of $F_{x,\mathrm{out}}$ and $S_{x,\mathrm{out}}$ as a function of $\overline{x}_{\mathrm{out}}$ for coherent light.} \label{f:fanoCOH}
\end{center}
\end{figure}

\begin{figure}[h]
\begin{center}{
 \includegraphics[angle=0,width=1\textwidth]{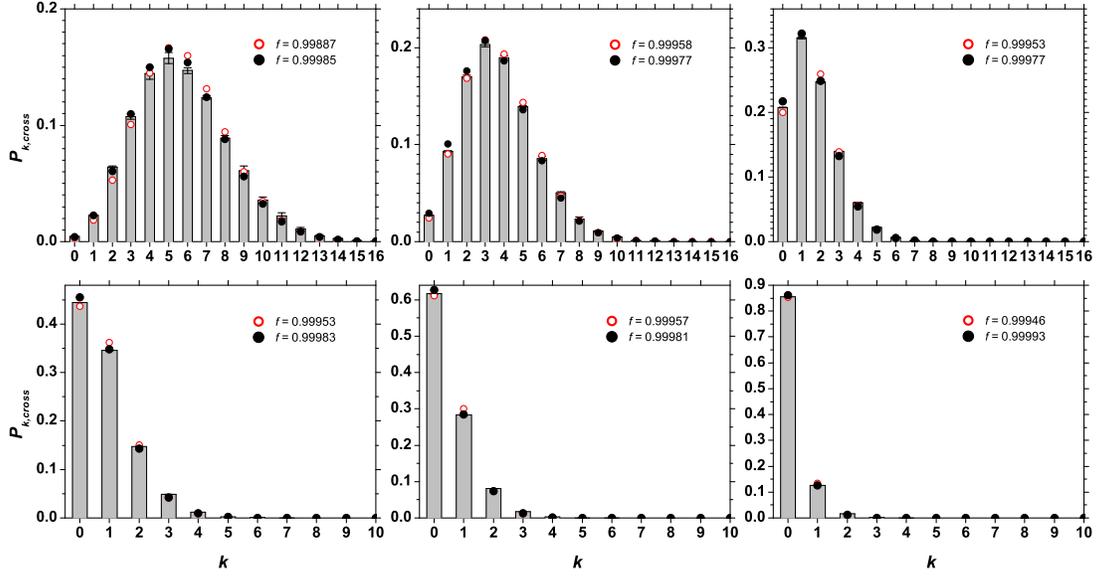} }
\caption{(Color online) Experimental $P_{k,\mathrm{cross}}$ distributions at different mean values (bars) and theoretical curves evaluated according to \emph{Model~I}: poissonian (white circles), poissonian modified by cross-talk effect (full circles). The corresponding fidelity values of the reconstruction are also shown.} \label{f:istoCOH}
\end{center}
\end{figure}

\begin{figure}[h]
\begin{center}{
 \includegraphics[angle=0,width=1\textwidth]{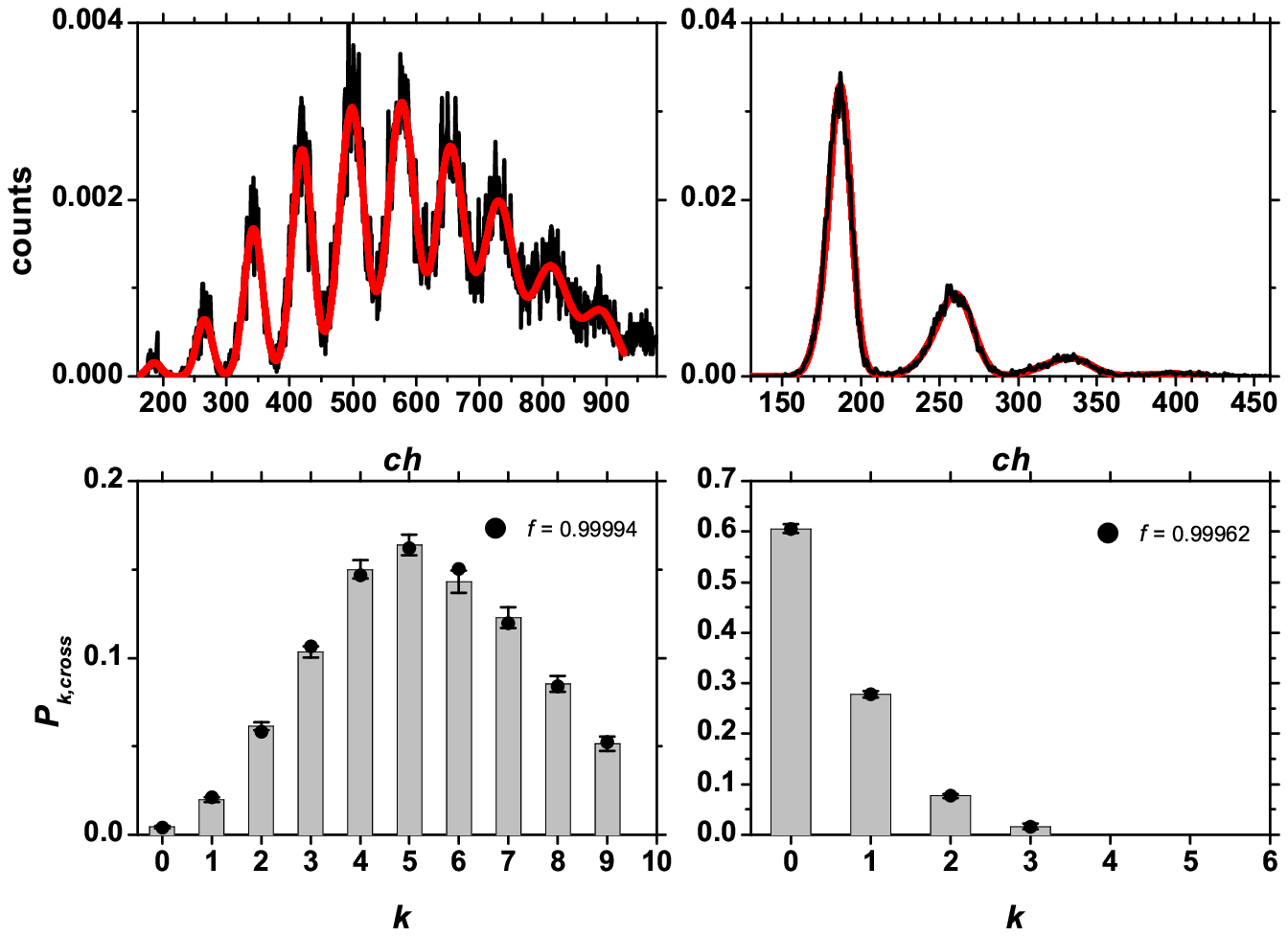} }
\caption{(Color online) Experimental results for \emph{Model~II} applied on two of the histograms acquired with coherent light. Upper row: result of the multi-peak fit procedure; lower row: fitted theoretical distributions. The corresponding fidelity values of the reconstruction are also shown.}\label{pic:poisson}
\end{center}
\end{figure}

First of all, we measured the coherent light emerging from the laser. We measured the
values of the output, $x$, at 20000 subsequent laser shots for 15 series, each one with a
different mean value $\overline{x}_\mathrm{out}$, set by rotating a polarizer (P in
Fig.~\ref{f:setup}) in front of the collection fiber. We then acquired a series of data
in the absence of light and set the zero at the mean value of the main peak of the
resulting histogram, in which the presence of dark counts and cross talk emerges as a
much lower separated peak. Following \emph{Model~I} we evaluated the experimental values
of $F_{x,\mathrm{out}}$ and $S_{x,\mathrm{out}}$ that are plotted in Fig.~\ref{f:fanoCOH}
as a function of $\overline{x}_\mathrm{out}$. We then fitted the data to straight lines
and obtained ($81.1\pm 0.2$)~ch, for $F_{x,\mathrm{out}}$, and ($6971\pm 57$)~ch$^2$, for
$S_{x,\mathrm{out}}$. These values were used to evaluate $\gamma$ and $\epsilon$ from
Eqs.~(\ref{eq:fanoCOH}) and (\ref{eq:skewCOH}). We obtained $\gamma = (75.4 \pm 1.3)$~ch
and $\epsilon = (0.039 \pm 0.009)$. The $x$-values were then divided by $\gamma$ and
re-binned in unitary bins \cite{JMO, Aa} to obtain the $P_{k,\mathrm{cross}}$
distribution of the actual light signal amplitude measured in the presence of dark-counts
and cross-talk. Note that due to the linearity of the detector the mean value of the
output can be directly obtained as
$\overline{k}_{\mathrm{cross}}=\overline{x}_{\mathrm{out}}/\gamma$, independent of the
shape of the distribution.\\
In Fig.~\ref{f:istoCOH} we plot as bars six different $P_{k,\mathrm{cross}}$
distributions at different mean values. Superimposed to the experimental values we plot
two theoretical curves, one is a poissonian (see Eq.~(\ref{eq:poiss})) having mean value
$\overline{k}_{\mathrm{cross}}$ (white circles), while the other (full circles) is
evaluated by including the cross-talk effect. We evaluate $P_{k,\mathrm{cross}}$ from
Eq.~(\ref{eq:crossDC}) in the case in which $P_{m,\mathrm{el+dc}}$ is poissonian and get:
\begin{equation}
 P_{k,\mathrm{cross}}=e^{-\overline{m}_{\mathrm{el+dc}}} (1-\epsilon)^{-k} \epsilon^k\
 _{p}F_{q}\left(1,-k;\frac{1}{2}-\frac{k}{2},1-\frac{m}{2};-\frac{(1-\epsilon)^2 \overline{m}_{\mathrm{el+dc}}}{4\epsilon}\right)\frac{\sin(k\pi)}{k\pi}\ , \label{eq:crossDCpoiss}
\end{equation}
where $\overline{m}_{\mathrm{el+dc}}=\overline{k}_{\mathrm{cross}}/(1+\epsilon)=
{\overline{x}}_{\mathrm{out}}/(\gamma (1+\epsilon))$. The theoretical distributions (full
circles in Fig.~\ref{f:istoCOH}) are evaluated by using the measured values of
$\overline{x}_{\mathrm{out}}$, $\gamma$ and $\epsilon$.\\
Note that measuring a coherent light with this method enables the simultaneous
characterization of the detector gain and of the contribution of cross-talk. The
comparison between the data and the theoretical functions can be estimated through the
evaluation of the fidelity
\begin{eqnarray}
 f=\sum_{k=0}^{m} \sqrt{P_{k,\mathrm{exp}} P_{k,\mathrm{theo}}}\ .\label{eq:fidel}
\end{eqnarray}
On the other hand, using the method of analysis of \emph{Model~II}, we could study each
one of the acquired histograms separately. As mentioned above, in the case of coherent
light, we have a theoretical fitting function with a total of three free parameters:
expectation value of light and DCR contribution $\overline{m}_{\mathrm{el+dc}}$, the
probability $\epsilon$ for an avalanche to trigger a second one and a global
normalization factor (up to three ``iterations''): in this case we are limited to spectra
with at least 4 resolved peaks.\\
In Fig.~\ref{pic:poisson} we show the results of this analysis for the multi-peak fit and
for the fit of the statistics of the avalanches using $\chi^2$ as an indicator of the
goodness of the fit result. The obtained $\epsilon$ values are compatible with what we
obtained with \emph{Model~I}. Also the gain values, evaluated as the peak-to-peak
distance, show a good agreement with the $\gamma$ values given by \emph{Model~I}: $\gamma
= (78.28 \pm 0.26)$~ch for spectrum in left panel and $\gamma = (72.61 \pm 0.19)$~ch for
spectrum in right panel.
\subsection{Multi-mode pseudo-thermal light}

\begin{figure}[h]
\begin{center}{
 \includegraphics[angle=0,width=0.7\textwidth]{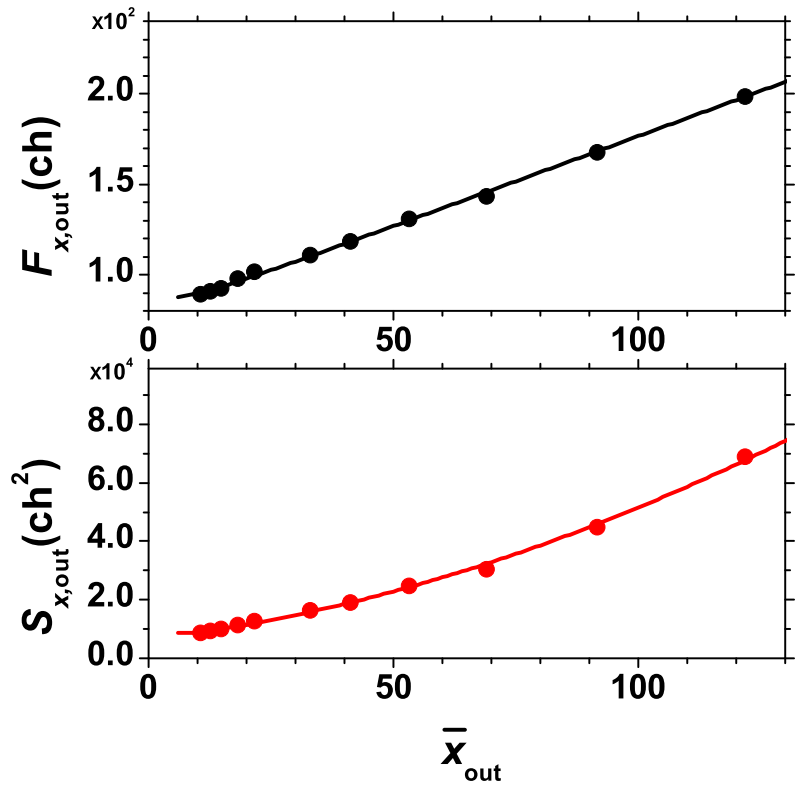} }
\caption{(Color online) Plot of $F_{x,\mathrm{out}}$ and $S_{x,\mathrm{out}}$ as a function of $\overline{x}_{\mathrm{out}}$ for pseudo-thermal light.} \label{f:fanoTHERM}
\end{center}
\end{figure}

\begin{figure}[h]
\begin{center}{
 \includegraphics[angle=0,width=1\textwidth]{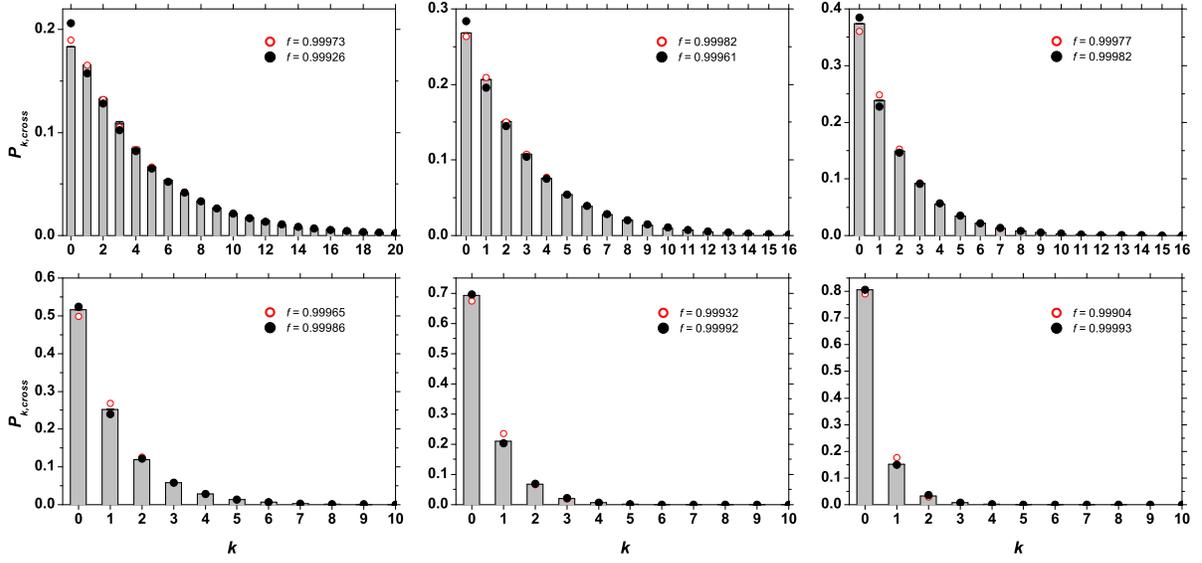} }
\caption{(Color online) Experimental $P_{k,\mathrm{cross}}$ distributions at different mean values (bars) and theoretical distributions evaluated according to \emph{Model~I}: thermal modified by dark count distribution (white circles), thermal modified by dark counts and cross-talk effect (full circles). The corresponding fidelity values of the reconstruction are also shown.} \label{f:istoTHERM}
\end{center}
\end{figure}

In order to obtain information on the contribution of dark-counts we have to measure a
different light statistics, whose shape is modified by the convolution with the
poissonian distribution for dark-counts. We thus produced a pseudo-thermal light field by
selecting with a small aperture ($\sim 150$~$\mu$m diameter) a region much smaller than
the coherence area of the speckle patterns produced by the rotating diffuser. We follow
the same procedure described for coherent light by measuring the values of the output,
$x$, at 50000 subsequent laser shots and at 10 different mean values, obtained by means
of a variable neutral-density filter (ND in Fig.~\ref{f:setup}). In
Fig.~\ref{f:fanoTHERM} we plot the experimental values of $F_{x,\mathrm{out}}$ and
$S_{x,\mathrm{out}}$ as a function of $\overline{x}_{\mathrm{out}}$. Along with the
experimental data we plot the fitting curves evaluated according to
Eqs.~(\ref{eq:FANOmultit}) and (\ref{eq:SKEWmultit}). To describe the fitting procedure
in detail we rewrite Eqs.~(\ref{eq:FANOmultit}) and (\ref{eq:SKEWmultit}) as the
\begin{eqnarray}
  F_{x,\mathrm{out}} &=& \left(1-\frac{\overline{x}_{\mathrm{dc}}}{\overline{x}_{\mathrm{out}}}\right)^2 \overline{x}_{\mathrm{out}}+ B\label{eq:SKEWmult}\\
  S_{x,\mathrm{out}} &=& A\left(1-\frac{\overline{x}_{\mathrm{dc}}}{\overline{x}_{\mathrm{out}}}\right)^3 \overline{x}_{\mathrm{out}}^2 +3B\left(1-\frac{\overline{x}_{\mathrm{dc}}}{\overline{x}_{\mathrm{out}}}\right)^2 \overline{x}_{\mathrm{out}} +C\ ,\label{eq:SKEWmult}
\end{eqnarray}
where we have set $\mu=1$. First of all we fitted the data to $F_{x,\mathrm{out}}$ and
obtained the values of $\overline{x}_{\mathrm{dc}} = (5.82028 \pm 1.34015)$ and $B =
(87.805 \pm 2.09009)$~ch. Then we fitted the data for $S_{x,\mathrm{out}}$ by
substituting the obtained values of $\overline{x}_{\mathrm{dc}}$ ch and $B$ to obtain $A
=(2.34754 \pm 0.091576)$ and $C=(8531.48 \pm 419.571)$~ch$^2$. These values are then used
to evaluate $\gamma$ and $\epsilon$ from Eqs.~(\ref{eq:FANOmultit}) and
(\ref{eq:SKEWmultit}). We obtained $\gamma = (74.2785 \pm 18.6017)$~ch and $\epsilon =
(0.100174 \pm 0.166894)$.
The $x$-values were then divided by $\gamma$ and re-binned in unitary bins \cite{JMO} to
obtain the $P_{k,\mathrm{cross}}$ distribution of the actual light signal amplitude
measured in the presence of dark-counts and cross-talk.
In Fig.~\ref{f:istoTHERM} we plot, as bars, six different $P_{k,\mathrm{cross}}$
distributions at different mean values. Superimposed to the experimental values we plot
two theoretical distributions: the first one (open circles) is evaluated by including the
contribution of dark-count that modifies the statistics of a single-mode thermal
distribution (see  Eq.~(\ref{eq:multit})) into $P_{m,\mathrm{el+dc}}$ according to
Eq.~(\ref{eq:phelDC}), which, in the present case, yields:
\begin{eqnarray}
 P_{m,\mathrm{el+dc}}&=&\sum_{k=0}^{m} P_{k,\mathrm{dc}} P_{m-k,\mathrm{el}} \nonumber\\
 &=&\frac{e^{-\overline{m}_{\mathrm{dc}}}}{(\mu-1)!} \left(1+\frac{\mu}{\overline{m}_{\mathrm{el}}}\right)^{-m} \left(1+\frac{\overline{m}_{\mathrm{el}}}{\mu}\right)^{-\mu}
  U\left[-m,1-m-\mu,\overline{m}_{\mathrm{dc}} \left(1+ \frac{\mu}{\overline{m}_{\mathrm{el}}}\right)\right]\ ,\label{eq:phelDC}
\end{eqnarray}
where $U(a,b,z)$ is the confluent hypergeometric function. The parameters are evaluated
as $\overline{m}_{\mathrm{dc}}= \overline{x}_{\mathrm{dc}}/(\gamma (1+\epsilon))$ and
$\overline{m}_{\mathrm{el}}=
(\overline{x}_{\mathrm{out}}-\overline{x}_{\mathrm{dc}})/(\gamma (1+\epsilon))$. The
second distribution (full circles) is evaluated from Eq.~(\ref{eq:crossDC}) to take into
account the cross-talk. Unfortunately, the calculation does not yield an easy analytical
result, and hence we evaluate it numerically. The values of the fidelity for the data in
Fig.~\ref{f:istoTHERM} improve when we take into account both dark-counts and cross-talk.

\begin{figure}[h]
\begin{center}{
 \includegraphics[angle=0,width=1\textwidth]{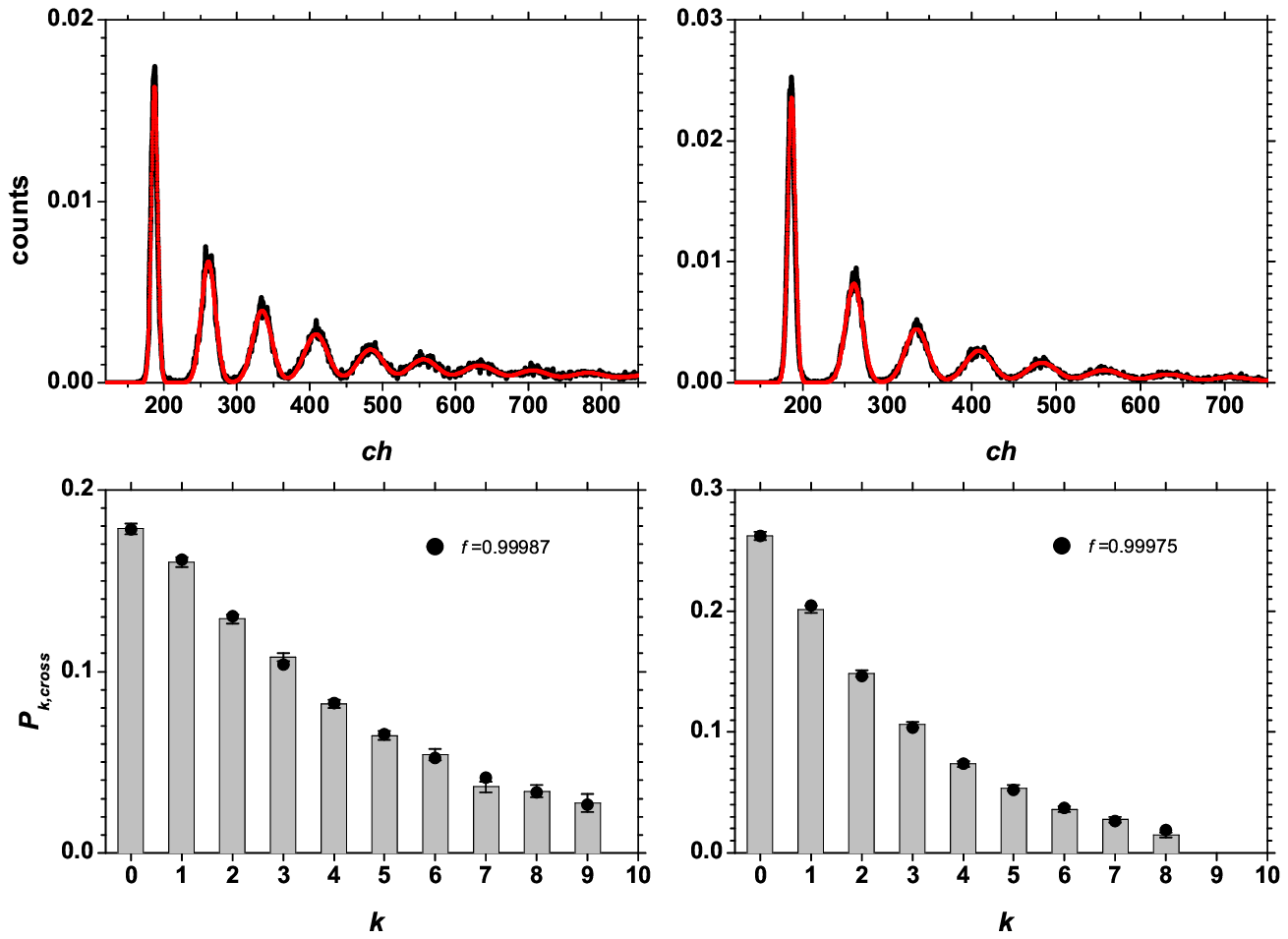} }
\caption{(Color online) Experimental results for \emph{Model~II} applied on two of the histograms acquired with thermal light. Upper row: result of the multi-peak fit procedure; lower row: fitted theoretical function. The corresponding fidelity values of the reconstruction are also shown.}\label{pic:thermic}
\end{center}
\end{figure}

Turning now to the other approach, we note how the number of fit parameters in this case
is enhanced: we now have the expectation value $\overline{m}_{\mathrm{el}}$ of avalanches
generated by detection, the expectation value $\overline{m}_{\mathrm{dc}}$ of DCR
contribution, the number of modes $\mu$, the probability $\epsilon$ of triggering a
cross-talk event (up to three iterations) and again a global normalization factor, for a
total of 5 fit parameters: obviously, this puts a severe limit on the
applicability of this method, needing at least 6 resolved peaks.\\
As it can be noted
from the fit results in Fig.~\ref{pic:thermic}, once again the results obtained  by using
\emph{Model~II} are compatible within errors with what we found by applying
\emph{Model~I}. However, whether the global fits present a very low $\chi^2$ value for
degree of freedom, the obtained fit parameters present high uncertainties, probably
indicating the presence of very high off-diagonal elements in the minimization matrix and
suggesting a strong correlation between the various parameters. This problem can be
avoided by fixing some of the fit parameters (such as DCR or cross-talk), by  retrieving
their value from an accurate direct measurement as that explained in
Section~\ref{sec:setup}.

\section{Discussion}\label{discussion}

\begin{figure}[h]
\begin{center}{
 \includegraphics[angle=0,width=1\textwidth]{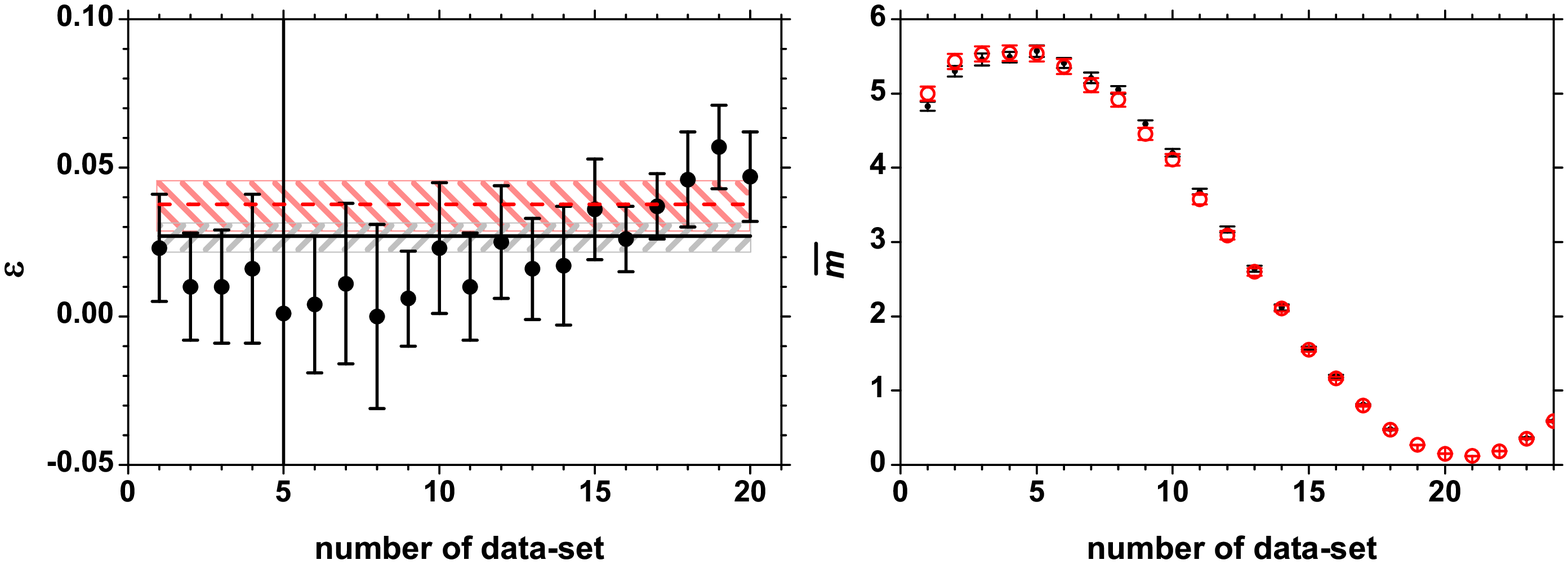} }
\caption{(Color online) Left panel: values of $\epsilon$
obtained by applying \emph{Model~II} to coherent light (full circles) and their weighted
average (full line). Dashed line: value of $\epsilon$ obtained by \emph{Model~I}.
Right panel: values of mean photon numbers evaluated by applying \emph{Model~II} (full circles) and by applying \emph{Model~I} (white circles).}\label{f:confrPOISS}
\end{center}
\end{figure}

We can compare the results of the two analysis methods on the same data-sets taken at
different intensities. In Fig.~\ref{f:confrPOISS} (a) we plot as full circles the values
of $\epsilon$ obtained by applying \emph{Model~II} to coherent light along with their
weighted average (full line). As a comparison, the value of $\epsilon$ obtained by
applying \emph{Model~I} is plotted as dashed line. In Fig.~\ref{f:confrPOISS} (b) we plot
as full circles the values of mean photon numbers evaluated for the same data as in panel
(a) by applying \emph{Model~II}. White circles represent the values of mean photon
numbers obtained by applying \emph{Model~I}. As we can see, the values are compatible
within errors.

\begin{figure}[h]
\begin{center}{
 \includegraphics[angle=0,width=1\textwidth]{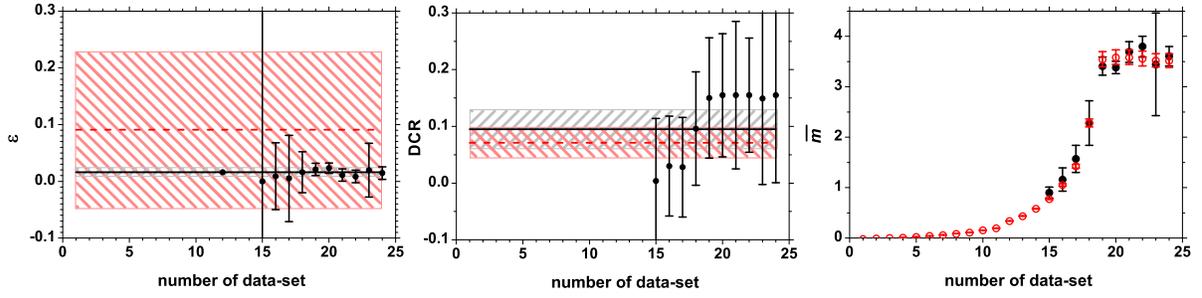} }
\caption{(Color online) Left panel: values of $\epsilon$
obtained by applying \emph{Model~II} to thermal light (full circles) and their weighted
average (full line). Dashed line: value of $\epsilon$ obtained by \emph{Model~I}.
Central panel: values of DCR evaluated for the same data as in the left panel by applying \emph{Model~II} (full circles) and their weighted average (full line). Dashed line: value of DCR obtained by \emph{Model~I}.
Right panel: values of mean photon numbers evaluated by applying \emph{Model~II} (full circles) and by applying \emph{Model~I} (white circles).}\label{f:confrTHERM}
\end{center}
\end{figure}

The same comparison for the measurements on thermal light is shown in
Fig.~\ref{f:confrTHERM}. Here we can see that the agreement is better for the mean values
of detected photons (panel (a)) and for the DCR (panel (b)), while the estimated values
of $\epsilon$ from the two \emph{Models} definitely disagree. This can be due to the
different approximations adopted by the two \emph{Models} (first order \emph{vs} third
order) that become relevant when measuring thermal light instead of coherent light. 

In Table~\ref{tab:compdcrxt} we summarize the results of the two \emph{Models}. We
demonstrated that both the \emph{Models} work in a self-consistent way, even if they have
two definitely different approaches. \emph{Model~I} does not need peak resolving
capability, but requires the acquisition of several histograms at varying $\eta$. Once
determined the parameters $\epsilon$ and DCR, all the data-sets in a series can be
analyzed, independent of the number of distinguishable peaks in the pulse-height
spectrum. \emph{Model~II} works analyzing each histogram independently, but, as
GM-avalanches distribution is obtained with a fit of the data, it requires at least a
number of resolved peaks greater than the number of free parameters. The fact that the
two \emph{Models} give very similar results for mean photon numbers is particularly
important as in most applications this is the only important parameter. Merging the two
\emph{Models} we can devise an optimal strategy based on a self-consistent calibration
performed by measuring a known light and analyzing the data with \emph{Model~I}: once
known $\epsilon$ and DCR, the determination of the mean photon number is independent of
the specific statistics of light. Hence the information on the mean photon number can be
obtained from each single measurement, even when the fitting procedure of \emph{Model~II}
cannot be applied.

\section{Conclusions}

We have developed a model to describe the operation of a multi-pixel detector for which
dark-count and cross-talk effects are non negligible and we have implemented two
different  procedures for recovering the values of dark-counts and cross-talk, both
implemented on measurements performed on the same light under characterization. We
demonstrated that \emph{Model~I} is self-consistent and does not need peak resolving
capability; on the other hand, it requires the acquisition of several histograms at
varying $\eta$. \emph{Model~II} works by analyzing each histogram independently, but
seems to be a less robust method in order to perform direct measurements of the detector
parameters and it is better performing if supported with direct measurement of some
relevant parameters. The results obtained by both methods show that the light statistics
can be reliably reconstructed in the case of coherent and thermal light.

\section*{Acknowledgments}
Part of the equipement used has been founded by the European Commission within the RAPSODI Project - COOP 32993.


\clearpage


\noindent Fig. 1. Left panel: Output of a SiPM Hamamatsu  MPPC S10362-11-100C detecting a
weak light field as displayed by a 100 MHZ oscilloscope. Right panel: histogram of the
corresponding spectrum.

\noindent Fig. 2. Threshold scan of the SiPM performed at room temperature with no
impinging light.

\noindent Fig. 3. (Color online) Experimental setup. Nd:YLF: laser source, P: polaroid,
ND: variable neutral density filter, SiPM: detector. The components in the dashed boxes
are inserted to produce the pseudo-thermal field.

\noindent Fig. 4. (Color online) Plot of $F_{x,\mathrm{out}}$ and $S_{x,\mathrm{out}}$ as
a function of $\overline{x}_{\mathrm{out}}$ for coherent light.

\noindent Fig. 5. (Color online) Experimental $P_{k,\mathrm{cross}}$ distributions at
different mean values (bars) and theoretical curves evaluated according to
\emph{Model~I}: poissonian (white circles), poissonian modified by cross-talk effect
(full circles). The corresponding fidelity values of the reconstruction are also shown.%

\noindent Fig. 6. (Color online) Experimental results for \emph{Model~II} applied on two
of the histograms acquired with coherent light. Upper row: result of the multi-peak fit
procedure; lower row: fitted theoretical distributions. The corresponding fidelity values
of the reconstruction are also shown.

\noindent Fig. 7. (Color online) Plot of $F_{x,\mathrm{out}}$ and $S_{x,\mathrm{out}}$ as
a function of $\overline{x}_{\mathrm{out}}$ for pseudo-thermal light.

\noindent Fig. 8. (Color online) Experimental $P_{k,\mathrm{cross}}$ distributions at
different mean values (bars) and theoretical distributions evaluated according to
\emph{Model~I}: thermal modified by dark count distribution (white circles), thermal
modified by dark counts and cross-talk effect (full circles). The corresponding fidelity
values of the reconstruction are also shown.

\noindent Fig. 9. (Color online) Experimental results for \emph{Model~II} applied on two
of the histograms acquired with thermal light. Upper row: result of the multi-peak fit
procedure; lower row: fitted theoretical function. The corresponding fidelity values of
the reconstruction are also shown.

\noindent Fig. 10. (Color online) Left panel: values of $\epsilon$ obtained by applying
\emph{Model~II} to coherent light (full circles) and their weighted average (full line).
Dashed line: value of $\epsilon$ obtained by \emph{Model~I}. Right panel: values of mean
photon numbers evaluated by applying \emph{Model~II} (full circles) and by applying
\emph{Model~I} (white circles).

\noindent Fig. 11. (Color online) Left panel: values of $\epsilon$ obtained by applying
\emph{Model~II} to thermal light (full circles) and their weighted average (full line).
Dashed line: value of $\epsilon$ obtained by \emph{Model~I}. Central panel: values of DCR
evaluated for the same data as in the left panel by applying \emph{Model~II} (full
circles) and their weighted average (full line). Dashed line: value of DCR obtained by
\emph{Model~I}. Right panel: values of mean photon numbers evaluated by applying
\emph{Model~II} (full circles) and by applying \emph{Model~I} (white circles).
%

\clearpage
\clearpage
\clearpage
\clearpage
\clearpage
\clearpage
\clearpage
\clearpage
\clearpage
\clearpage
\clearpage
\begin{table}[h]
\begin{center}
\begin{tabular}{lcc}
\hline
\hline
      & Hamamatsu MPPC S10362-11-100C \\
\hline
     Number of Diodes: & 100\\
     Area: & 1~mm $\times$ 1~mm\\
     Diode dimension: & 100~$\mu$m $\times$ 100 $\mu$m \\
     Breakdown Voltage: & 69.23~V \\
     dark-count Rate: & 540~kHz  at 70 V \\
     Optical Crosstalk: & 25~$\%$ at 70 V \\
     Gain: & $3.3~\cdot~10^6$ at 70 V\\
     PDE (green): & 15~$\%$ at 70 V \\
\hline
\hline
\end{tabular}
\end{center}
\caption{\label{tab:MPPC} Main characteristics of the SiPM (Hamamatsu, model MPPC
S10362-11-100C). The data refer to room temperature.}
\end{table}
\clearpage
\begin{table}[tb]
\begin{tabular}{l c|c|c|c|c|}
\cline{3-6}
 & & \multicolumn{2}{|c|}{Poissonian} & \multicolumn{2}{|c|}{Pseudo-thermal}\\ \cline{3-6}
 & & \emph{Model~I} & \emph{Model~II} & \emph{Model~I} & \emph{Model~II}\\
\hline
\hline
\vline &DCR &- &- & 0.071 $\pm$ 0.027& 0.094 $\pm$ 0.035  \\
\vline &$\epsilon$ & 0.038 $\pm$ 0.008 & 0.027 $\pm$ 0.005 &  0.091 $\pm$ 0.138 & 0.035 $\pm$ 0.004  \\
\hline
\hline
\end{tabular}
\caption{\label{tab:compdcrxt} Comparison between the global DCR and cross-talk values
obtained with \emph{Model~I} and weighted average of the values obtained with
\emph{Model~II}.}
\end{table}
\end{document}